\documentclass{emulateapj}

\usepackage{epsfig}
\usepackage{natbib}

\usepackage{apjfonts}
\usepackage{xspace}
\usepackage{hhline}
\usepackage{amsmath}
\usepackage{multirow}

\makeatletter
B  {\def\@captype{table}}
  {}

\makeatother

\usepackage{psfig}
\usepackage{graphicx}
\usepackage{times}
\usepackage{xspace}
\def\ni{\noindent}

\def\Halpha{H$\alpha$}
\def\about{$\sim$}
\def\arcsec{$\,^{\prime\prime}$~}
\def\arcmin{$\,^\prime$~}
\def\deg{$^{\circ}$}
\def\erg/cm2sec{ergs~cm$^{-2}$~s$^{-1}$}  
\def\ergcm2{ergs~cm$^{-2}$}  
  
\def\mdot{$\dot{m}$~}  
\def\X{$\times$}
\def\Fx{F$_X$~}

\def\Av{A$_V$~}
\def\Fr{F$_R$~}
\def\Ar{A$_R$~}
\def\FxFv{{F$_X$/{F$_V$}}}
\def\FxFr{{F$_X$/{F$_R$}}}
\def\Lx{L$_x$~}

\def\Msun{$M_\odot$}

\def\ncv{n$_{cv}$~}
\def\pc3{pc$^{-3}$~}
\def\cm3{cm$^{-3}$~}

\newcommand{\lsim }{{\lower0.8ex\hbox{$\buildrel <\over\sim$}}}
\newcommand{\gsim }{{\lower0.8ex\hbox{$\buildrel >\over\sim$}}}
\newcommand{\atoms}{\ifmmode{\rm ~atoms~cm^{-2}} \else ~atoms cm$^{-2}$\fi}
\newcommand{\cmsq}{\ifmmode{\rm ~cm^{-2}} \else cm$^{-2}$\fi}
\newcommand{\nh}{{\ifmmode{\rm N_{H}} \else N$_{H}$\fi}}
\newcommand{\nhgal}{\ifmmode{ N_{H}^{Gal}} \else N$_{H}^{Gal}$\fi}
\newcommand{\nhintr}{\ifmmode{ N_{H}^{intr}} \else N$_{H}^{intr}$\fi}
\newcommand{\nhtot}{\ifmmode{ N_{H}^{tot}} \else N$_{H}^{tot}$\fi}
\newcommand{\meangamma}{\ifmmode{\langle\Gamma\rangle} \else $\langle\Gamma\rang
le$\fi}
\newcommand{\fx}{\ifmmode L_x \else $~f_x$\fi}
\newcommand{\fxfopt}{\ifmmode \frac{f_x}{f_{opt}} \else $\frac{f_x}{f_{opt}}$\fi
}
\newcommand{\logfx}{\ifmmode{\rm log}~f_x \else log$~f_x$\fi}
\newcommand{\logfxfopt}{\ifmmode{\rm log}\,(\frac{f_x}{f_{opt}}) \else
${\rm log}\,(\frac{f_x}{f_{opt}})$ \fi}
\newcommand{\lopt}{\ifmmode L_{opt} \else $~L_{opt}$\fi}
\newcommand{\loglopt}{\ifmmode{\rm log}~L_{opt} \else log$~L_{opt}$\fi}
\newcommand{\lx}{\ifmmode L_x \else $~L_x$\fi}
\newcommand{\loglx}{\ifmmode{\rm log}~L_x \else log$~L_x$\fi}
\newcommand{\aox}{\ifmmode{\alpha_{ox}} \else $\alpha_{ox}$\fi} 
\newcommand{\aro}{\ifmmode{\alpha_{ro}} \else $\alpha_{ro}$\fi} 
\newcommand{\luv}{\ifmmode L_{uv} \else $~L_{uv}$\fi}
\newcommand{\logluv}{\ifmmode{\rm log}\,L_{uv} \else log$\,L_{uv}$\fi}
\newcommand{\bmv}{\ifmmode{(B-V)} \else $(B-V)$ \fi}
\newcommand{\ebmv}{\ifmmode{\rm E}_{B-V} \else E$_{B-V}$\fi}
\newcommand{\nufnu}{\ifmmode \nu f_{\nu} \else$\nu f_{\nu}$\fi}
\newcommand{\fnu}{\ifmmode f_{\nu} \else$f_{\nu}$\fi}
\newcommand{\fcgs}{\ifmmode {\rm erg~cm}^{-2}~{\rm s}^{-1}\else
erg~cm$^{-2}$~s$^{-1}$\fi} 
\newcommand{\lcgs}{\ifmmode erg~~s^{-1}\else erg~s$^{-1}$\fi}
\newcommand{\flamcgs}{\ifmmode erg~cm^{-2}~s^{-1}~Hz^{-1} \else erg~cm$^{-2}$~s$
^{-1}~$\AA$^{-1}$\fi}
\newcommand{\fnucgs}{\ifmmode erg~cm^{-2}~s^{-1}~Hz^{-1}\else erg~cm$^{-2}$~s$^{
-1}$~Hz$^{-1}$\fi}
\newcommand{\lnucgs}{\ifmmode erg~s^{-1}~Hz^{-1}\else erg~s$^{-1}$~Hz$^{-1}$\fi}
\newcommand{\kms}{\ifmmode~{\rm km~s}^{-1}\else ~km~s$^{-1}~$\fi}
\newcommand{\mone}{\ifmmode ^{-1}\else$^{-1}$\fi}
\newcommand{\mtwo}{\ifmmode ^{-2}\else$^{-2}$\fi}
\newcommand{\degs}{\ifmmode ^{\circ}\else$^{\circ}$\fi}
\newcommand{\mv}{\ifmmode {m_{V}}\else${m_{V}}$\fi}
\newcommand{\Mv}{\ifmmode {M_{V}}\else${M_{V}}$\fi}
\newcommand{\msun}{\ifmmode {M_{\odot}}\else${M_{\odot}}$\fi}
\newcommand{\rsun}{\ifmmode {R_{\odot}}\else${R_{\odot}}$\fi}
\newcommand{\lsun}{\ifmmode {L_{\odot}}\else${L_{\odot}}$\fi}

\slugcomment{To appear in ApJ}

\begin{document}
\pagestyle{plain}
\pagenumbering{arabic}

\title{Chandra Multiwavelength Plane (ChaMPlane) Survey: an Introduction}

\author{J. E. Grindlay, J. Hong, P. Zhao, S. Laycock, M. van den Berg, 
X. Koenig and E.M. Schlegel\altaffilmark{1} \\
H.N. Cohn, P.M. Lugger and A.B. Rogel\altaffilmark{2}}

\altaffiltext{1}{Harvard-Smithsonian Center for Astrophysics, 60 Garden St,
Cambridge, MA 02138; josh@cfa.harvard.edu}

\altaffiltext{2}{Department of Astronomy, Indiana 
University, 727 E. Third Street, Bloomington, IN 47405}

\begin{abstract}
We introduce  the Chandra Multiwavelength Plane (ChaMPlane) Survey, 
designed to measure or constrain the populations of low-luminosity 
(\Lx \gsim10$^{31}$ \lcgs) accreting white dwarfs, neutron stars 
and stellar mass black holes in the Galactic Plane and Bulge.  
ChaMPlane  incorporates two surveys, X-ray (Chandra) 
and optical (NOAO 4m-Mosaic 
imaging), and a followup spectroscopy and IR identification program. 
The survey has now extended through 
the first 6 years of Chandra data using serendipitous 
sources detected in 105 distinct ACIS-I and -S fields observed 
in 154 pointings and covered by 65 deep Mosaic images in 
V, R, I, and \Halpha. ChaMPlane incorporates fields with galactic 
latitude $|b|$ \lsim12\deg~ and selected to be devoid 
of bright point or diffuse sources, with  exposure 
time \gsim20ksec, and (where possible) minimum \nh. 
We describe the scientific goals and introduce the X-ray and 
optical/IR processing and databases. 
We derive preliminary constraints on the space density or 
luminosity function of cataclysmic variables 
from the X-ray/optical data for 14 fields in the Galactic Anticenter. 
The lack of ChaMPlane CVs in these Anticenter fields suggests their 
space density is \about3\X~ below the value 
(3 \X~ 10$^{-5}$ pc$^{-3}$) found  
for the solar neighborhood by previous X-ray surveys. 
Companion papers describe the X-ray and optical processing 
in detail, optical spectroscopy of ChaMPlane sources in 
selected Anticenter fields  and IR imaging results for the 
Galactic Center field. An Appendix introduces the ChaMPlane Virtual 
Observatory (VO) for online access to the X-ray and optical images 
and source catalogs for ready display and further analysis.
\end{abstract}


\maketitle

\section{Introduction}

The  Chandra Multiwavelength Plane (ChaMPlane) Survey is 
a comprehensive effort to analyze systematically and 
archive the results from deep (\gsim20-100ksec) Chandra pointings near 
the Galactic Plane ($|b|$ \lsim 12\deg). The primary science goals 
of ChaMPlane (described below) are to measure or constrain the 
low-luminosity (\lsim10$^{34}$ \lcgs) accretion-powered 
X-ray sources in the Galaxy: 
accreting white dwarfs in cataclysmic variables (CVs), 
neutron stars and black holes in quiescent low mass X-ray binaries 
(qLMXBs) and low-luminosity high mass X-ray binaries (HMXBs), 
primarily Be-HMXBs. ChaMPlane is also enabling the first 
high-resolution search for isolated stellar mass black holes accreting 
from giant molecular clouds (GMCs). A key secondary goal is the 
study of stellar coronal emission across the H-R diagram and 
in a range of Galactic environments: from the Anticenter disk to 
the Galactic Bulge. The survey will produce 
the most comprehensive database of coronal X-ray sources (stars) 
yet achieved, ultimately expanding the ROSAT samples 
(e.g. Schmitt et al 2004 and references therein) by 
\about1-2 orders of magnitude for stars in the plane 
and (far) beyond the solar neighborhood. 
While the primary goals of ChaMPlane are to survey and 
study point sources, the ChaMPlane X-ray and optical image  
archive also contain considerable rich diffuse emission which can be 
accessed for further study. 

ChaMPlane is also a 
major optical imaging (V, R, I and \Halpha) followup survey 
conducted to identify optical counterparts of the X-ray survey sources.  
We have used the Mosaic camera on the CTIO and KPNO 4m telescopes  
in a 5-year {\it Long Term Survey Program} granted to ChaMPlane 
(in 2000) by NOAO. The images and photometry from this 
survey, separately processed and archived at both CfA and 
NOAO, constitute some of the deepest \Halpha~ images of the 
Plane in the 65 Mosaic fields (36\arcmin \X 36\arcmin) obtained in 
our NOAO survey. The ChaMPlane database at CfA links the X-ray and 
optical images and, when complete, will produce a legacy 
database for the distribution and 
nature of faint X-ray sources in the Galactic Plane. With initial 
analysis on 14 Anticenter fields now nearly complete, the 
fraction of sources with optical counterparts is \about50\%.

ChaMPlane is deeper (given diffuse 
emission in many fields) and of course has  
much higher angular resolution, permitting optical/IR identifications, 
than the XMM Galactic Plane 
survey (Motch et al 2003), ROSAT all sky survey (Voges et al 1999) 
or the Einstein Galactic Plane 
Survey (Hertz and Grindlay 1984; Hertz et al 1990).  
Given that Chandra is the premier high-resolution 
X-ray imager for at least the next 
decade, it is important to conduct the systematic 
analysis of the (many) Chandra Galactic 
fields  to create the  database needed for developing 
a statistical understanding of the accreting binary (white 
dwarf, neutron star and black hole) as well as stellar coronal 
X-ray content of the Galaxy. 
An initial description of ChaMPlane and 
preliminary results for the derived source number vs. flux 
(logN-logS) distributions and source types  
was given for several Galactic Bulge fields by 
Grindlay et al (2003) along with an initial description 
of the optical survey (Zhao et al 2003). 

The X-ray processing is described in detail in the accompanying 
paper by Hong et al (2005), 
which presents X-ray results for the initial 14   
Galactic Anticenter fields (90\deg~ \lsim $l$ \lsim 270\deg). 
The optical survey and 
photometry pipeline processing for ChaMPlane 
is described in the accompanying 
paper by Zhao et al (2005). A broad overview of both the 
X-ray and optical survey status, as well as portals 
to the archived data and source catalogs, is available 
from the ChaMPlane website 
\footnote{http://hea-www.harvard.edu/ChaMPlane/}. 
ChaMPlane also incorporates an extensive optical spectroscopy 
followup program (at WIYN, MMT, CTIO and Magellan) and  
infrared (IR) imaging and photometry and ultimately spectroscopy. 
Details are given in sections 5.2 and 5.3.

The ChaMPlane X-ray and optical surveys 
are now building, and releasing, significant 
X-ray and optical database archives of processed images and derived 
X-ray fluxes and colors and optical photometry. The optical 
database also includes photometry (V,R,I and \Halpha) 
and astrometry results for the 
full stellar content of all Mosaic fields acquired. 
The images and derived catalogs 
are released as initial science papers are submitted. 
We have developed tools for easy web 
access to the X-ray and optical images 
and derived products in the database. Examples of these tools for 
X-ray and optical overlays are given in the Appendix. 

We first summarize the key science objectives of ChaMPlane and 
the criteria for selection of Chandra pointings for the survey. We 
then describe the X-ray processing and survey coverage  
obtained and then the optical-IR surveys and data products.  
Finally, we derive initial constraints on the CV density and/or
luminosity function from X-ray/optical results for the 14 fields 
processed for the Anticenter. In the Appendix we 
provide an overview of the X-ray and optical 
database content and online display and analysis tools now 
available from the ChaMPlane website.

\section{Primary Scientific Objectives: 
Accreting Compact Objects in the Galaxy}

ChaMPlane uses Chandra data from 
ACIS\footnote{http://cxc.harvard.edu/proposer/POG/html/ACIS.html}  
pointings, primarily 
ACIS-I, as this provides a larger contiguous field of view (16\arcmin 
\X 16\arcmin), with exposure times of \gsim20 ksec. A 
typical ChaMPlane field (see Galactic distribution of ChaMPlane 
fields and histograms of key parameters in 
Figures 1 and 2) has an absorption column with log (\nh) \about21.7. 
Thus using PIMMS\footnote{http://asc.harvard.edu/toolkit/pimms.jsp}  
and an assumed power law spectrum 
with index 1.7, the limiting flux sensitivity for a 10 ct source 
in a ``typical'' 30 ksec observation  
is \Fx(0.5-8 keV) = 5.7 \X 10$^{-15}$ \fcgs (unabsorbed), yielding a 
distance limit for a source with \Lx(0.5-8 keV) = 10$^{31}$ \lcgs of 
d$_{31-typ}$ \about3.9kpc. For the deepest ChaMPlane exposures 
of typically 100ksec, and the same \nh, this increases to 
d$_{31-max}$ \about7.1kpc, and for those extreme fields (e.g. Baade's 
Window) with both deep exposure and minimal log (\nh) = 21.3, the 
limiting flux is \Fx = 1.3 \X 10$^{-15}$ \fcgs (unabsorbed) for 
``limiting'' distance d$_{31-lim}$ \about8.2 kpc. The 
corresponding minimum luminosity for a more heavily 
reddened field with log (\nh) = 22.3 at d = 8kpc (e.g. 
much of the Galactic Center region) and a 100 ksec exposure 
with 10ct detection limit is \Lx(0.5-8 keV) = 2.5 \X 10$^{31}$ \lcgs.  
These values for limiting distance and \Lx guide the 
principal scientific objectives for ChaMPlane.

\smallskip

{\it CV Number Density:} Our first objective is to measure, 
or significantly limit, the number, space density and luminosities 
of cataclysmic variables (CVs) in the Galactic Plane.  
Current estimates of CV space density as \about10$^{-5}$pc$^{-3}$ 
(Patterson 1998) are largely based on the small number
(\about300) of CVs detected  
from optical (variability or color) surveys and may be  
uncertain by at least a factor of 10 (Warner 1995).    
ROSAT surveys have shown that indeed the optical 
surveys are (very) incomplete; 
a significant number (\gsim80) of CVs have now been identified in
ROSAT fields, mostly with magnetic CVs (Schwope et al 2002 
and references therein).  
Determining the true CV space density and Galactic 
population is important since this must connect to problems as 
fundamental as the rate of novae in the Galaxy 
and the origin of SN Ia systems (e.g. Townsley and Bildsten 2005) 
as well as the population of LMXBs and CVs in the Bulge 
that originated in globular clusters, either by ejection or 
cluster disruption  (Grindlay 1985). 
Estimates of the CV space density from the Einstein Galactic 
Plane Survey (Hertz et al 1990) and ROSAT Bright Star survey 
(Schwope et al 2002) have both  given \ncv \about 
3 \X 10$^{-5}$ \pc3. Although the Einstein/ROSAT CVs have reached 
farther than the optical surveys, 
none is more distant than \about1200pc, 
and very little is known about the CV space density in the 
Bulge or radial distribution in the Disk. 
Extrapolating the local \ncv estimates, a 
typical  ACIS-I ChaMPlane field might then contain 
\about0.2-2 CVs. We derive  predictions below (Table 2), 
and compare them with preliminary results for 14 Anticenter fields. 
The effects of \nh, assumed spatial distribution, CV X-ray 
luminosity function and X-ray/optical flux distributions 
make such predictions uncertain. No CVs 
have been confirmed spectroscopically  
as counterparts to Chandra sources in these fields, whereas \about1 
might be expected given our discovery of 5 optically-selected CVs 
in the \about5X larger area of the corresponding optical Mosaic 
images. As also pointed out below, the fraction of optical 
counterparts from the Mosaic imaging that have been spectroscopically 
identified in these Anticenter fields is only \about30\%, so definitive CV 
measures are still to be derived. 

\smallskip

\ni{\it Black Hole vs. Neutron Star LMXBs:}  
The space density of low mass X-ray binaries, with 
either black hole (BH) or neutron star (NS) primaries, 
is even more uncertain. The total number of BH systems 
in the Galaxy is probably  \gsim10$^3$, as
estimated from the numbers and recurrence times of 
BH X-ray novae (Tanaka and Lewin 1995). 
The number of LMXBs in quiescence, or qLMXBs, containing 
NSs is even more uncertain yet a large 
population is indicated by the much larger population of millisecond
pulsars (MSPs) in the Galaxy (cf. Yi and Grindlay 1998) and 
population of faint transients in the Bulge (cf. Heise et al 1998).  
Since both qLMXBs and CVs have X-ray luminosities typically
\about10$^{30.5-32.5}$ \lcgs, they are best discovered by a deep X-ray 
survey with flux sensitivity sufficient to detect 
these across the Galaxy and positional resolution sufficient to 
allow unambiguous optical or IR identification. 

\smallskip

\ni
{\it Galactic Center and Bulge-Cusp sources:} 
The population of faint sources in the central cusp about SgrA* 
(Muno et al 2003) may be dominated by wind-fed HMXBs, most 
likely quiescent Be-HMXBs. Alternatively as also  
proposed by Muno et al, they may be relatively luminous 
CVs, most likely intermediate polars. Our 
deep IR imaging (JHKBr$\gamma$) of the SgrA* field 
(Laycock et al 2005) shows that the bulk of these sources 
do not have high mass companions with spectral types 
B0 or B1 as expected for Be-HMXBs, and thus may be magnetic CVs. The 
nature and number of the Bulge-Cusp sources, vs. the Galactic Bulge and 
disk sources generally, are major objectives of the ChaMPlane survey. 
This has motivated our targeted program of deep pointings on three 
low-extinction windows progressively closer to the Galactic Center 
(see Fig. 1); these are reported separately.
\smallskip

\ni
{\it Isolated BHs in GMCs:} Whereas stellar mass BHs (sBHs) are now 
readily detected in binaries (usually as transients in low mass X-ray 
binaries) as hard X-ray sources, isolated sBHs should be much more 
numerous. Agol and Kamionkowski (2002) estimate there may be 
\about10$^{8-9}$ sBHs in the Galaxy distributed in a disk distribution 
with scale height \about250pc. If these are born with kick velocities 
much less than those derived for NSs, as is plausible for SN II core 
collapse models, they should be detectable by their Bondi-Hoyle 
accretion when passing 
through dense molecular clouds and GMCs. Assuming only \about1\% gas-capture 
efficiency of Bondi-Hoyle accretion (Perna et al 2003), and an 
additional radiative 
efficiency factor $\epsilon \sim10^{-4}$ (allowing for an ADAF-like flow), 
the (hard) X-ray luminosity expected can still be appreciable. For a 
10\Msun~ BH with velocity V = 10 km s$^{-1}$ passing through a cold GMC 
(with sound speed c$_s <<$ V, so that (V$^2$ + c$_s^2$))$^{3/2}$ 
\about V$^3$)  with density n = 10$^4$ \cm3, the accretion luminosity is 
\medskip

\Lx \about 3.4 \X $10^{31} \epsilon_{-4} n_4 M_{10}^2/V_{10}^3$ \lcgs

\medskip

\ni
where the $\epsilon$, n, M and V parameters are scaled to the values  
given above. Given the sensitivities above, this \Lx could be detected 
with ChaMPlane in a ``typical'' Galactic Bulge field, with \nh~ \about 
2 \X 10$^{22}$ \cmsq, in a 100 ksec pointing. The challenge, of course, 
would be to identify (in the IR, given the extinction in the Bulge 
and the GMC, which alone gives \nh~ \about 3 \X 10$^{23}$ \cmsq) the nature 
of such a source. With no binary companion, it should have 
an X-ray/IR flux ratio (unabsorbed) Fx/F$_{IR}$ \gsim1 
(since the unabsorbed \Lx/L$_{IR}$ >1 for any re-processed thermal IR) 
vs. Fx/F$_{IR}$ \about 0.1 for the X-ray to K band flux ratios for 
the BH-LMXBs GROJ0422+32 or A0620-00. However, setting lower limits 
on K magnitudes will be difficult because stellar crowding in Bulge 
fields limits K \lsim16 (Laycock et al 2005) and thus isolated 
BHs in GMCs may be best identified by high resolution 
(NICMOS or AO) IR observations 
of highly self-absorbed ChaMPlane sources, or  
in shallower but wide surveys of relatively nearby 
GMCs with less stellar crowding. 

\smallskip

\ni
{\it Be X-ray Binaries:} Be stars in binaries with 
NS companions (only, thus far) are perhaps the most numerous of the high 
mass X-ray binaries (HMXBs) in the Galaxy. ChaMPlane is sensitive 
to the underlying population of quiescent Be-HMXB (qBe-HMXB)   
systems; those in outburst are obvious as bright transients. 
From the values for \FxFv~ given for qBe-HMXBs by 
van Paradijs and McClintock (1995), and converting to conventional 
broad-band values, we derive log(\FxFv) \about -2.5 $\pm$0.9 for 
qBe-HMXBs which have mean  \Lx = 10$^{33.3 \pm1.2}$ \lcgs. 
Isolated Be stars, on the other hand, should have 
\Lx \about10$^{-7}$L$_{bol}$ \about10$^{29.7}$ \lcgs (for a B0V star) 
and thus log(\FxFv) \about -5.5 and so be readily distinguished.    
Be-HMXB systems with BHs as the compact object  
are expected from binary and stellar evolution, 
and are another key goal of the ChaMPlane survey. 
Neutron star accretors, 
as found thus far, can be determined by X-ray pulsation 
analysis (all Be-HMXBs known thus far are accreting pulsars) 
although limited counting statistics makes this test possible only 
for the brightest sources.  

\begin{figure}
\resizebox{\hsize}{!}
{\includegraphics[bb=155 90 470 710,angle=-90]{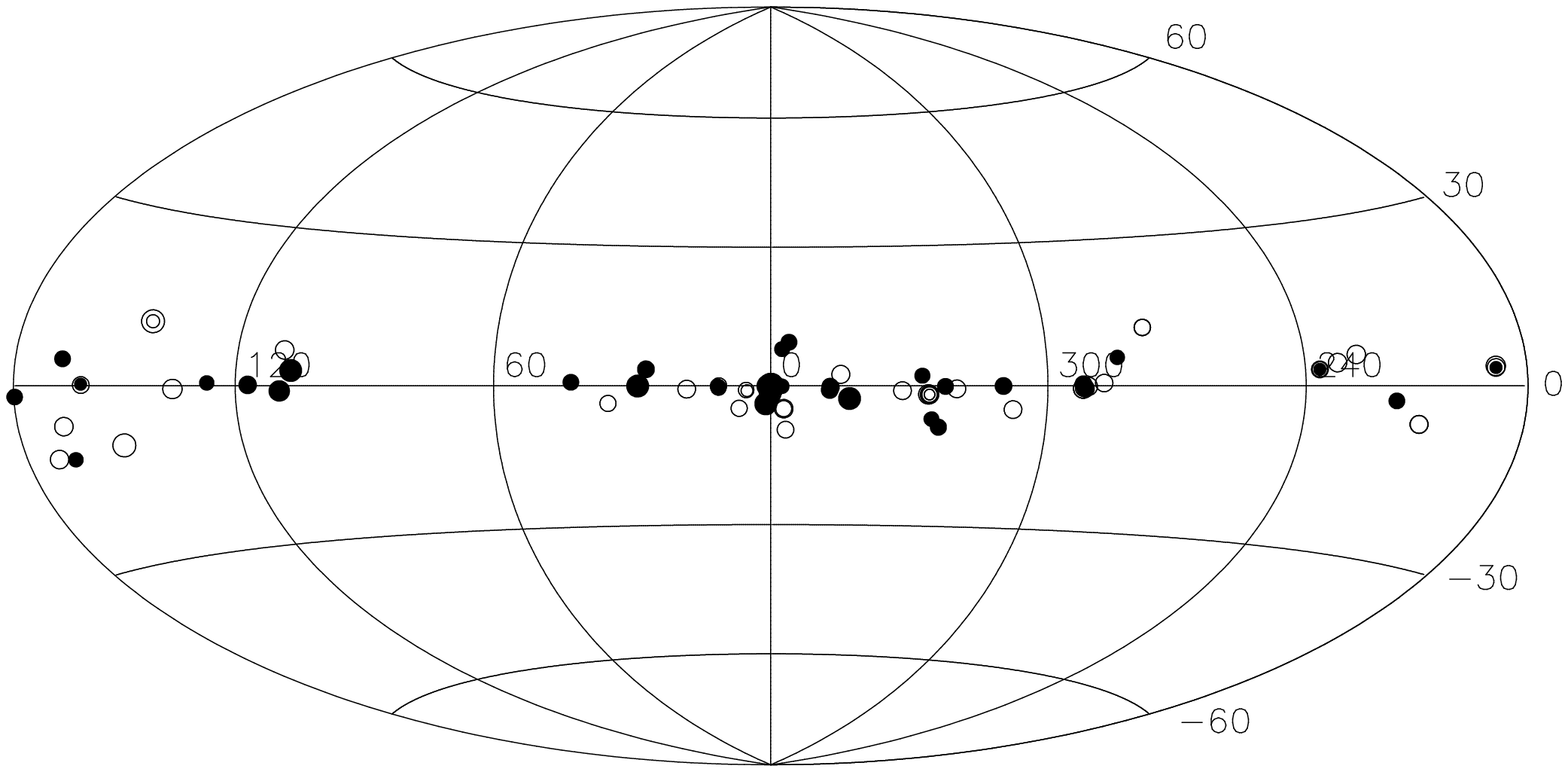}}
{\includegraphics[bb=56 25 402 195,width=8.8cm,height=4.cm]{f1b.ps}}

\caption{{\it Top:} Galactic distribution of 154 observations 
of 105 distinct ChaMPlane fields with separations 
at least \gsim4\arcmin through Chandra cycle 6. Plotted 
positions are ACIS-I (solid) vs. -S (open), with circle 
diameters scaled to exposure time. 
{\it Bottom:} Positions of 40 distinct ChaMPlane fields,  
with 62 observations, in the Galactic Center region. Included 
are our deep (100ksec) obervations of 3 low extinction fields: Baade's
Window at $l, b$ = 1\deg, -3.6\deg,  
``Stanek's Window'' at $l,b$ = 0.2\deg,-2.1\deg~ to be followed by a 
``Limiting Window'' (in Chandra cycle 6) at $l,b$ = 0.1\deg, -1.4\deg.}
\label{Fig1}
\end{figure}

\smallskip

\ni{\it Stellar Coronal Luminosity Functions:} 
Most soft Galactic Plane Chandra sources are identified 
(from positional matches) as coronal emission from
stars, primarily K, M and dMe stars, but extending throughout the 
entire H-R diagram. With typically \gsim15 coronal sources per field, 
the entire survey (\about100 fields) should yield \gsim1500 stars, with 
photometric spectral types constrained by color-color analysis 
(cf. Zhao et al 2005) and reddening constrained by X-ray colors 
(cf. Hong et al 2005) or actually measured for a 
significant fraction from followup optical spectroscopy. 
Since ChaMPlane covers a wide 
range of the Galaxy, from fields 
in the Anticenter to the Bulge, these will allow entirely new 
constraints on stellar coronal emission over a range of 
metallicity and stellar populations not possible before with 
ROSAT (e.g. Schmitt et al 2004).

\section{Selection Criteria for Survey Fields}

ChaMPlane uses both ACIS-I and 
ACIS-S Chandra pointings, which are imaging only (i.e. no grating 
data), without sub-array or continuous clocking data mode (which 
restrict the field of view) and 
which meet the criteria: 
\begin{enumerate}
\item preferably ACIS-I, for the larger field of 
view nearly on-axis this enables;
\item exposure times nominally \gsim20ksec (though some, like the 
Galactic Center survey (Wang et al 2002), are included with 12ksec 
exposures) and galactic latitude $|b|$ \lsim12\deg;   
\item do not contain bright point sources or large/bright diffuse
X-ray emission or extended clusters, which would systematically 
both limit sensitivity and contaminate survey fields with "targets", and; 
\item have minimal \nh, though this choice is rarely possible to make. 
\end{enumerate}

\ni
These are primary criteria, in approximate priority order, used 
to select ChaMPlane fields and, in turn, to choose coordinates for 
the deep optical imaging (Mosaic) for our NOAO-ChaMPlane survey . 
In a few cases, we find a short observation (5-10ksec) will be 
included within a Mosaic field previously acquired. These are 
included to help in extending the survey areal coverage and thus 
sensitivity to bright sources in the Plane with lower space density. 

The fields selected for cycles 1-6   
are given on the ChaMPlane website, with their full parameters, 
and are shown schematically in Fig. 1. The distributions of 
exposure time, \nh, Galactic longitude and latitude of the 
105 distinct ChaMPlane fields chosen thus far 
are shown in Fig. 2. In defining ``distinct'' fields, we have 
required pointing positions to be different by at least 4\arcmin, 
since this is a characteristic scale for appreciable degradation 
of on-axis sensitivity and also the scale of half an ACIS CCD. 
The additional 49 ChaMPlane pointings (in the current total 
of 154) thus represent at least partial overlap fields, which 
allow significant new opportunities for time variability studies 
as part of the survey. 
ChaMPlane also  includes the 30 fields observed by 
Wang et al (2002) for their survey of the Galactic Center region.  
These are counted as distinct (since their offset is 12\arcmin) 
although in fact they overlap by 8\arcmin and thus allow 
variability searches for bright sources moderately far off-axis. 

An overview of the current multiple 
exposure coverage, in galactic longitude vs. temporal coverage, is 
provided in Fig. 3. A ``typical'' ChaMPlane field is \about30 ksec 
(Fig. 2) exposure and contains  \about70  
sources in each ACIS-I (or ACIS-S, scaled for area) 
field, with each source located to \lsim1\arcsec precision in the 
central \about6\arcmin (diameter) field of view (FoV) 
and to within \lsim2\arcsec even near the edge of the FoV. 

\begin{figure}
\resizebox{\hsize}{!}
{\includegraphics[angle=0]{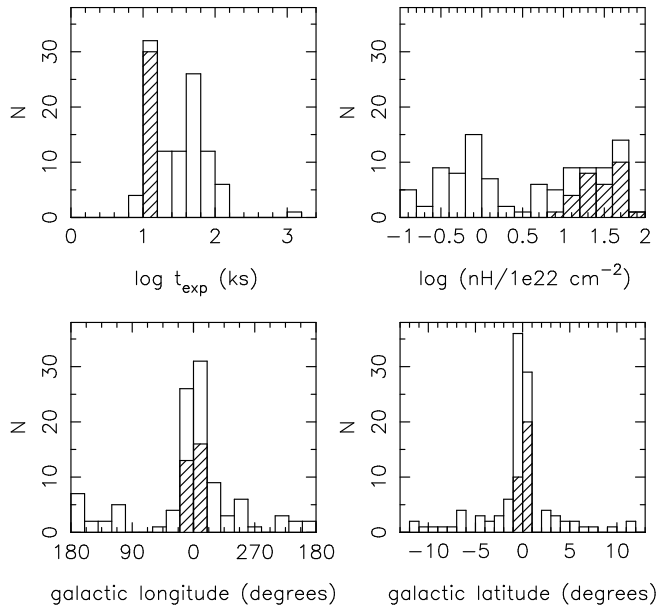}}
\caption{Distributions of exposure time, \nh, and galactic 
coordinates $l, b$ for 105 distinct ChaMPlane fields currently selected 
for ChaMPlane. Cross-hatched bins are the 12 ksec Galactic Center
survey fields of Wang et al (2002). The extreme exposure bin 
of \about1Msec represents the current total of all observations 
on SgrA*.}
\label{Fig2}
\end{figure}

\begin{figure}
\resizebox{\hsize}{!}
{\includegraphics[angle=0]{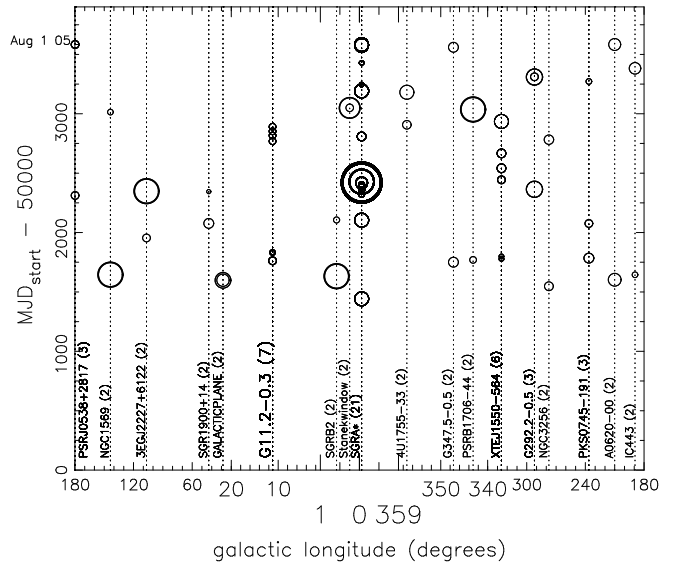}}
\caption{Distributions of ChaMPlane fields with multiple exposures. 
Times (MJD) of individual exposures for ChaMPlane fields with 
overlapping (\lsim4\arcmin) pointing positions are plotted vs. 
galactic longitude, with target name labelled. Exposure times 
for each exposure are proportional to the diameter of the marker 
circles. Three cycle 6 observations not yet carried out have been given the 
arbitrary date of Aug. 1, 2005, for plotting purposes.}
\label{Fig3}
\end{figure}

\section{X-ray Survey Processing and Data Products}

\begin{table}
\small

\begin{center}
\caption{Current vs. expected ChaMPlane coverage (ACIS-I vs. -S)}

\begin{tabular}{lrrrrr}
\hline\hline
& & No. of & sources & Total & area \\ 
Obs. type & No. Fields & lev 2 & lev 3 & (deg$^2$) & (deg$^2$) \\
ACIS-I & 63 & 5299 & 3998 & 4.92 & 2.43 \\
ACIS-S & 27 & 1746 & 1001 & 1.98 & 0.53 \\
Current Total  & 90 & 7045 & 4999 & 6.90 & 2.96 \\
\about Final Total & 105 & 8219 & 5832 & 8.05 & 3.45 \\
\hline
\end{tabular}
\end{center}
Source totals of distinct ChaMPlane fields 
processed to date (Aug. 2005) for level 2 vs. level 3 
sources (see text and Hong et al 2005 for definitions) 
and respective total survey areas covered. Level 3 sources 
and survey areas are a subset of the corresponding Level 2 
values. The final row gives the approximate extrapolated 
totals expected for the current total of 105 distinct 
ChaMPlane fields.
\end{table}

ChaMPlane fields are processed with  a uniform detection 
and processing pipeline, XPIPE, based on the script 
developed by Kim et al (2004) for the Chandra high-latitude 
survey, ChaMP. ChaMPlane incorporates significant additional 
processing in a Post XPIPE Processor (PXP) as described in detail 
by Hong et al (2005). The entire processing system 
incorporates a number of new features of (potential) 
general interest to the Chandra community, including 
a very thorough treatment of X-ray source location 
confidence radii (e.g. 95\% confidence) vs. off-axis 
radii and X-ray to optical boresighting, with total 
uncertainties (see also Zhao et al 2005). 
Source detection and verification (see 
Hong et al 2005 for details) produce increasingly 
restrictive source catalogs: level 1, 2 and 3. Level 2 sources 
are the nominal primary database for ChaMPlane and are derived 
from detections on the 4 ACIS-I CCDs (CCDs 0, 1, 2, 3) for ACIS-I 
observations or the  2 ACIS-S CCDs (CCDs 6, 7) and two contiguous 
ACIS-I CCDs (CCDs 2, 3) for ACIS-S observations if they were all 
enabled. Some ACIS-S observations have fewer CCDs 
enabled so that the areal coverage for level 2 ACIS-S sources 
is not always 4 CCDs per observation. 
Level 3 sources are closer to on-axis and thus have smaller 
error circles and higher sensitivity. For ACIS-I observations, they 
are confined to within 400\arcsec from the Chandra aimpoint; for 
ACIS-S observations they are confined to just the backside illuminated 
primary S3 chip (CCD 7). 

Some 90 distinct fields (of 105 total) have been currently 
processed for the ChaMPlane survey. 
Source numbers and areal coverage for these fields are 
given in Table 1. The ACIS-I vs. -S areal coverage does not 
scale exactly with the number of fields for level 2 sources 
because of the differing number of CCDs enabled for -S 
observations, as noted above. A total of 119 observations (of the 154 
total) have been processed, so that the total source numbers 
are currently 10,283 (lev 2) and 7518 (lev 3). Comparison with 
Table 1 totals for distinct sources then shows that approximately  
3238 (lev 2) and 2519 (lev 3) sources are multiple observations, 
though in fact some of these are new sources. 

Whereas XPIPE processing and 
wavdetect\footnote{http://cxc.harvard.edu/ciao/ahelp/wavdetect.html} 
source detection are done in the same broad band, Bx(0.3-8.0 keV), 
as ChaMP (Kim et al 2004), the larger \nh~ expected for ChaMPlane 
sources dictates our choice of ``conventional'' bands 
Bc(0.5-8.0 keV), Sc(0.5-2.0 keV) and Hc(2.0-8.0 keV) 
which are extracted in the PXP script from the counts in 
the wavdetect source regions and used for broad-band analysis 
such as logN-logS. X-ray colors and rough spectral 
classification are not done using conventional hardness ratios 
and color-color analysis, but rather more generally with 
quantile analysis (Hong, Schlegel and Grindlay 2004). Quantile 
color color diagrams (QCCDs)  
enable more meaningful X-ray color analysis on weak sources. 
LogN-logS and quantile analysis results for 14 Anticenter 
fields are given by Hong et al (2005). 

Output data products (source positions and uncertainties, fluxes 
in conventional bands for X-ray/optical 
flux ratios (e.g., \FxFv, \FxFr, etc.) are stored 
in a searchable database accessible through the ChaMPlane website 
for which search and display tools have 
been developed (see Appendix). 
Images and data products for the Anticenter 
observations (14 fields) described by Hong et al (2005)  
are now incorporated in this ChaMPlane database and are 
available from the ChaMPlane website (see Appendix).

\section{Optical-IR Survey and Data Products}

Here we provide an 
overview of the three principal elements of our source 
identification program: Mosaic imaging and photometry, 
spectroscopic followup, and IR imaging (and spectroscopic) 
followup. 

\subsection{Photometry from ChaMPlane Mosaic images}
Our NOAO optical survey program obtained 65 deep 
stacked Mosaic images in each of 4 filters 
that include all the 105 Chandra fields 
shown in Fig. 1. The details of our Mosaic photometry analysis 
are given by Zhao et al (2005). 

Accretion sources may be  identified by their {\it ubiquitous
\Halpha~ emission:} most CVs (except dwarf novae in outburst), and all
qLMXBs (and bright LMXBs) show \Halpha~ in emission 
(see below for EW(\Halpha) statistics) from their
accretion disks or accretion columns (in the case of 
Polars). Be/X-ray binaries, and isolated Be stellar 
X-ray sources are also (by definition) 
\Halpha~ emitters. Consequently, 
the ChaMPlane NOAO Survey program was designed as a wide-field deep 
\Halpha~ imaging survey, with annual allocations of 5 nights of 
deep V, R, \Halpha, I (to R\about24) imaging time on the 
CTIO-4m and 1-2 nights on the KPNO-4m telescopes. 
We use the NOAO Mosaic cameras, which provide 36\arcmin \X 36\arcmin
fields and therefore a factor of \about5 more  
optical survey coverage than the 16\arcmin \X 16\arcmin~ ACIS-I field. 
The larger Mosaic imaging thus provides a deep optical 
comparison survey for CVs and emission line objects. 
\Halpha--R vs. R color 
magnitude diagrams identify \Halpha~ counterpart candidates for 
ChaMPlane sources (Grindlay et al 2003, Zhao et al 2003). 

Color-color diagrams [(V-R) vs. (R-I)] are constructed 
for each field as an additional aid for initial source 
classification using X-ray constraints on source 
class (e.g. stellar coronal vs. accretion): objects can 
be de-reddened using 
estimates of \nh~ from QCCDs. When optical spectra 
are eventually obtained, we employ a technique 
to constrain source distances and \nh~ (and thus X-ray 
luminosities) by using the photometry in combination 
with spectroscopy and the reddening models of Drimmel 
et al (2003); details are given in Koenig et al (2005, 
in preparation). 

Our requirement to image CVs and qLMXBs, with typical 
absolute magnitudes \Mv \about 7, at 8 kpc and with (at least) 
Av = 2.5, leads to the desired magnitude limit of R=24 for the survey. 
Since the primary discriminant is \Halpha--R, we seek 5\% photometry 
at R=24 and 10\% photometry in \Halpha~ as well as V and I,
requiring a total \about4h integration per field with the 
NOAO 4m telescopes and reasonable (\about1\arcsec) seeing. 

\subsection{Spectroscopic followup program}
Spectroscopic followup is sought on all optical IDs 
that can be reached with WIYN-Hydra or 
the MMT-Hectospec for northern fields and either CTIO-4m-Hydra 
or Magellan-IMACS for southern fields. 
Our strategy is to obtain spectra for all Chandra 
source optical counterparts with R \lsim22 
(as dictated by WIYN, MMT, CTIO-4m and Magellan), with highest priority 
for those with \Halpha~ excess as given by the Mosaic photometry. 
We use  ``extra'' fibers/slits for \Halpha~ emission 
candidates identified with Mosaic 
in the surrounding larger field. We have obtained spectra of 
bright (V \lsim17) objects with FAST 
on the FLWO-1.5m telescope. 

Initial results for spectroscopic followup with WIYN-Hydra for 
13 ChaMPlane fields, including 11 of the 14 Anticenter fields 
presented here, have been reported 
by Rogel et al (2005). At least 5 CVs have 
been discovered in the surrounding 
Mosaic fields, but none inside the Chandra ACIS fields (see 
below). Followup spectroscopy is most incomplete for Galactic Bulge fields 
where we expect the largest accretion source content. 
We have obtained some spectra (Magellan-LDSS2, Magellan-IMACS 
and CTIO-Hydra) in 16 out of 25 distinct ChaMPlane 
fields in the Galactic Bulge ($\mid l\mid$ \lsim 20\deg), but 
coverage is still very incomplete. Initial results for LDSS2 
spectra are presented by Koenig et al (2005, in preparation).  

\subsection{IR photometry program}
Given the large excess of 
sources near the plane of the  
Galactic Bulge, and (apart from the low extinction ``windows'')   
given the high extinction which precludes all but sources 
in the foreground \lsim3 kpc from optical identifications and spectra, we 
have implemented an extensive IR photometry program to complement 
the Mosaic optical imaging for these Bulge fields. Initial results 
from a 10\arcmin \X 10\arcmin mosaic of IR-imaging (J, H, K, Br$\gamma$) 
around SgrA* we have obtained with the 
newly commissioned PANIC intrument on Magellan are presented by 
Laycock et al (2005).

\section{Preliminary Results for CVs in the Anticenter}

Although analyis is still underway for the full complement of 
22 Anticenter fields, X-ray processing results for 14 of them 
are summarized here (Table 2) for comparison with initial 
preditions for the possible contribution of CVs. Details 
of the X-ray processing, including logN-logS analysis 
and example QCCDs for comparison with known sources are 
given in Hong et al (2005). Optical photometry results for one 
field, GROJ0422+32 (at $l, b$ = 165.88\deg, -11.91\deg),  
are included in Zhao et al (2005), and 
will be reported separately for the remaining Anticenter fields. 
Table 2 gives the Chandra Observation ID (for easy reference; see 
Appendix), target, galactic coordinates,  exposure time achieved, 
\nh~ value for absorption through the 
full Galaxy at each position as derived from the extinction map of 
Schlegel, Finkbeiner and Davis (1998), approximate flux 
sensitivity limit in the Hc band (2.0 - 8 keV) 
achieved for an assumed power law spectrum with photon 
index 1.7 and the number of Chandra sources detected in each field 
for level 2 vs. level 3 (Hong et al 2005) processing of ACIS-I 
(all 4 CCDs) vs. ACIS-S (single S7 CCD), respectively. The final 
three columns of Table 2 are discussed in section 6.2. Although 
the nominal ChaMPlane survey sources are level 2 for both ACIS-I 
and -S observations, in the initial analysis presented here we 
use the more restrictive level 3 cut for ACIS-S observations 
for the simplification of not having mixed CCD types (i.e. back- vs. 
front-side illuminated chips for -S observations). 

\subsection{Source Properties}
Although 367 of the 631 level 2 (ACIS-I) and level 3 (ACIS-S) 
Anticenter sources have optical counterparts, 
only 4 of these were identified, securely, with 
apparent \Halpha~ emission 
objects with (\Halpha--R) \lsim -0.3: the two qLMXB black hole 
binaries, GROJ0422+32 and A0620-00, a QSO at z = 4.25 (with Ly$\alpha$ 
redshifted into the \Halpha~ filter) as a counterpart of a 
source in the G116.9+0.2 field, 
and a second source in this field which could be affected 
by nebular \Halpha~ emission. A number of marginally 
bright \Halpha~ sources with (\Halpha--R) \lsim -0.2, corresponding 
to equivalent widths (see Zhao et al 2005 for the (\Halpha--R) vs. 
EW(\Halpha) calibration) of EW(\Halpha) \about 18\AA\footnote{all EW 
values refer to {\it emission} lines; we drop the negative sign}, 
are probable counterparts. WIYN Hydra spectra of some of these objects 
(Rogel et al 2005) reveal most of these to be dMe stars. Spectra 
have been obtained for 141 (including two BH-LMXB targets)  
of the 279 ChaMPlane 
sources with optical matches in 11 of these 14 fields 
and have shown 59 to be coronal emission from stars, 
19 to be AGN, 5 possible AGN, and 58 to be unidentified 
due to, in most cases, insufficient S/N. 
Given variable observing conditions, not all of the  
fields were observed with photometric errors as small as 
shown (Zhao et al 2005) for GROJ0422+32. Even in this 
field, with low \nh~ and relatively high latitude, only 
40 of the 62 ChaMPlane sources have optical counterparts 
brighter than the magnitude limit R \about24.5. For the X-ray flux 
limit F$_{Hc}$ \gsim 1.1 \X 10$^{-14}$ \fcgs for this field, 
the corresponding unabsorbed flux limit 
is log (\FxFr) \gsim 1, implying the still-fainter unidentified 
sources are very likely accretion-powered sources -- most 
likely AGN, although this value is at the upper envelope 
(e.g. Brandt and Hasinger 2005) unless they are significantly 
self-absorbed.   

Indeed, as shown by Hong et al (2005), logN-logS analysis in the Hc band 
shows the sources in these 14 fields are apparently dominated by 
AGN since the ChaMPlane number counts are consistent 
with the high-latitude Chandra counts derived by ChaMP (Kim et al 2004).   
Although optical spectroscopy (Rogel et al 2005) shows that stars 
are a significant fraction (\about42\%) of ChaMPlane counterparts 
for which spectra have been obtained in Anticenter fields, these 
59 stars are only 23\% of the optical matches in the 11 fields for 
which spectroscopy was obtained and are also only 13\% of the total 
503 sources in these 11 fields. Allowing for the incomplete spectral 
coverage thus far in these fields (only 141 of 279 optical 
IDs in the 11 fields), 
the fraction of stellar coronal source IDs could be doubled to \about25\% 
(or \about 10-20 sources per field). Since stellar coronal sources are also 
primarily detected in the Sc (0.5-2 keV) band for 
which the logN-logS results of Hong et al (2005) show that the 
Anticenter sources may exceed the AGN background, the 
results are self-consistent. The stellar vs. AGN contributions are 
also constrained by the X-ray to optical flux ratio (\FxFr) 
distributions. 

\medskip

\begin{table*}
\small

\begin{center}
\caption{Initial CV constraints for 14 ChaMPlane Anticenter fields}

\begin{tabular}{rlccccccccccc}
\hline\hline
ObsID & Target & $l$   & $b$  & Exp.  & N$_{22}$    & log(F$_{Hc}$) &Nx &
d$_{30}$-d$_{31}$ & {CV$_{30}$-CV$_{31}$} & {R$_{30}$-R$_{31}$} &
{C$_{30}$-C$_{31}$} & {ID$_{30}$-ID$_{31}$}\\ 
& & (\deg~)&(\deg~)&(ksec )&(\cmsq )&    &   &          (kpc)
&       & (mag)  & & \\   
\hline
2787 & PSR J2229+6114 & 106.65 & 2.95  & 92-I  & 0.99  & -14.40 & 85 & 
1.5-4.7 & 0.6-14.5  &21-28 & 0.52-0.22 & 0.3-3.2 \\
755 & B2224+65 & 108.64 & 6.85  & 48-S  & 0.42  & -14.37 & 39 & 
1.4-4.5 &0.1-2.0 & 20-26 & 0.72-0.52 & 0.1-1.0 \\
2810 & G116.9+0.2 & 116.94 & 0.18  & 49-I  & 0.46  & -14.25 & 94 & 
1.3-4.0 &0.4-13.4 & 20-26 & 0.72-0.52 & 0.3-6.9 \\
2802 & G127.1+0.5 & 127.11 & 0.54  & 19-I  & 0.91  & -13.90 & 41 & 
0.8-2.6 &0.1-3.9  & 19-26 & 0.86-0.52 & 0.1-2.0 \\
782 & NGC1569 & 143.68 & 11.24  & 93-S  & 0.40  & -14.55 & 53 & 
1.8-5.6 &0.2-1.4& 21-26 & 0.52-0.52 & 0.1-0.7 \\
650 & GK Per & 150.96  &-10.10  & 90-S  & 0.20  & -14.44 & 35 & 
1.6-5.0 &0.1-1.4& 20-25 & 0.72-0.72 & 0.1-1.0 \\
2218 & 3C 129 & 160.43 & 0.14  & 30-S  & 0.63  & -13.93 & 14 & 
0.9-2.7 & 0-1.1 & 19-25 & 0.86-0.72 & 0.0-0.8 \\
676 & GROJ0422+32 & 165.88  &-11.91  & 19-I  & 0.20  & -13.95 & 62 & 
0.9-2.8 &0.1-1.7 & 19-24 & 0.86-1.0 & 0.1-1.7 \\
2803 & G166.0+4.2 & 166.13 & 4.34  & 29-I  & 0.36  & -14.08 & 52 & 
1.0-3.3 &0.2-4.7  & 19-25 & 0.86-0.72 & 0.2-3.4 \\
829 & 3C123 & 170.58  &-11.66  & 46-S  & 0.57  & -14.14 & 29 & 
1.1-3.5 &0-0.7  & 20-26 & 0.72-0.52 & 0.0-0.3 \\
2796 & PSR J0538+2817 & 179.72 &-1.68   & 19-S  & 0.82  & -13.91 & 12 & 
0.9-2.7 &0-0.9  & 19-26 & 0.86-0.52 & 0.0-0.5 \\
95 & A0620-00 & 209.96 &-6.54   & 41-S  & 0.28  & -14.22 & 29 & 
1.2-3.9 &0.1-1.4 & 19-25 & 0.86-0.72 & 0.1-1.0 \\
2553 & Maddalena's Cloud & 216.73 &-2.60 & 25-I & 0.97 & -13.96 & 55 & 
0.9-2.8 &0.1-3.9  & 19-26 & 0.86-0.52 & 0.1-2.0 \\
2545 & M1-16 & 226.80 & 5.63   & 49-S  & 0.13  & -14.36 & 31 & 
1.4-4.5 &0.1-2.2& 19-25 & 0.86-0.72 & 0.1-1.6 \\
Totals& & & & & & & 631 & & 2.3-53.1 & &  & 1.7-26.3 \\ 
\hline
\end{tabular}
\end{center}
Column headings are: Chandra observation ID (ObsID); target name; 
galactic coordinates $l, b$; achieved 
exposure time (ksec) in 
ACIS-I or -S; maximum column density \nh~ {\it in units of 10$^{22}$ \cmsq} 
from Schlegel et al (1998); 
3$\sigma$ detection sensitivity limit (unabsorbed) in Hc (2.0-8.0 keV) 
band, averaged over field;  number of 
sources detected in Bx (0.3-8.0 keV) band over 4 ACIS-I or 1 ACIS-S chips; 
range of limiting detection distance d$_{30}$ and d$_{31}$ 
for \Lx = 10$^{30-31}$ \lcgs; corresponding expected number of X-ray 
CVs detectable (CV$_{30}$ and CV$_{31}$) in effective volume 
out to distances d$_{30}$ and d$_{31}$ after 
scaling from \ncv = 3 \X 10$^{-5}$ \pc3 in the solar neighborhood; 
corresponding range of R magnitudes expected {\it at the typical 
extremes of \FxFr}, where R$_{30}$ is 
R magnitude expected for \Lx = 10$^{30}$ \lcgs and 
log(\FxFr) = -1 and R$_{31}$ is for \Lx = 10$^{31}$ \lcgs and 
log(\FxFr) = +1. R magnitudes have been absorbed for the NH expected 
out to d$_{30}$ and d$_{31}$ as shown in Fig. 6 for each field. 
Thus the low \Lx \about10$^{30}$ \lcgs CVs are expected primarily between 
magnitudes R$_{30}$ and R$_{30}$ + 5, and typical CVs with \Lx
\about10$^{31}$ \lcgs therefore between magnitudes R$_{31}$ and 
R$_{31}$ - 5. C$_{30}$ and C$_{31}$ are the cumulative fractions of 
CVs at the corresponding magnitude limits that would be detectable 
with R \lsim23 (see Fig. 7), and ID$_{30}$ and ID$_{31}$ are the 
corresponding numbers of optically identified CVs expected. 

\end{table*}

As an initial check on the CV population and contribution 
to the ChaMPlane Anticenter sources vs. AGN and stars, we 
plot in Figure 4 the \FxFr~ distributions for the Sc(0.5-2.0keV) 
band for the optically identified sources, along with the 
spectroscopically confirmed stars and AGN (Rogel et al 2005) for 
a subset of these with \gsim3$\sigma$ flux measurements in 
the Sc band. We use here the Sc band for ready comparison with 
both the ROSAT CVs (Verbunt et al 1997 and Schwope et al 2002) 
and the Chandra high latitude survey, ChaMP, population of AGN 
(and foreground stars) in this same Sc band. The total ChaMPlane 
optical source distribution of 324 of the 367 IDs for which \FxFr~ 
values can be derived (43 have poorly determined R magnitudes) 
is double peaked. From comparison with the spectroscopic 
samples, composed predominantly of stars (left peak) and
AGN (right peak), the 131 sources with \gsim3$\sigma$ flux 
measurements in the Sc (0.5-2.0 keV) band are consistent with 
being \about32\% stars and 68\% AGN (CVs and other accretion-powered 
galactic sources are also included, predominantly, in the hard peak).  

The 117 ChaMP AGN (taken from Green et al 2004) 
are overplotted in Figure 4 (dashed) and have been reddened to 
the ChaMPlane \nh values by dividing the ChaMP sample 
into 14 groups with numbers of sources proportional 
to the relative source numbers Nx in Table 2 for the Anticenter 
fields. Each group is then reddened in both \Fx and \Fr for 
the \nh~ in that field, and a corrected \FxFr~ derived. The peak of the 
reddened ChaMP distribution matches closely the right peak of the 
ChaMPlane distribution, further demonstrating that most are AGN. 
However so does (approximately) the peak of the ROSAT CV distribution, 
which is UNreddened (most ROSAT CVs are nearby and have minimal \nh)
and lines up approximately with the peak of the (unreddened) ``raw'' 
ChaMP AGN. The CVs may differ from the AGN in having a tail to lower 
values of \FxFr~ as well as a broader peak to higher \FxFr~ values.  
Thus CVs, particularly magnetic CVs but also 
qLMXBs (with both NS and BH primaries), with 
their flat Bremsstrahlung continua, are often harder than AGN 
in their \FxFr~ flux ratios. 

Moving to the harder Hc band also increases 
the contrast (separation) between the CVs (and qLMXBs) vs. stars. 
In Figure 5 we plot the log(\FxFr) distributions for the 68 
observed and unabsorbed (for the full-plane \nh) 
sources with \gsim3$\sigma$ fluxes in the Hc band. The 5 source  
``peaks'' at log(\FxFr) \about -2 and -3 (raw vs. unabsorbed, 
respectively) include 3 stars confirmed spectroscopically so 
that, as expected, coronal stellar sources are almost gone 
in the Hc band.  Thus, in Table 2, and the discussion that 
follows for initial constraints on CVs, we consider the 
Hc band only. 

\begin{figure}
\resizebox{\hsize}{!}
{\includegraphics[angle=0]{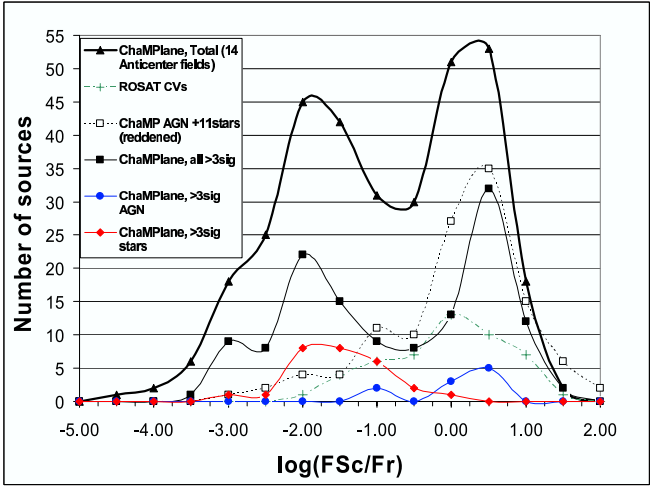}}
\caption{log(\FxFr) distributions (F(Sc) vs. R band)  
for 324 ChaMPlane Anticenter sources (heavy curve; filled triangles), 
131 with well-measured (\gsim3$\sigma$) F(Sc) fluxes (solid  
squares), and their decomposition into 27 spectroscopically 
confirmed stars (solid diamonds) and 
10 AGN (filled circles), with the remainder either unobserved or 
unclassified due to limited S/N. Comparison distributions are 
plotted for 49 ROSAT CVs (dot-dashed curve; + symbols) and for 
117 ChaMP high latitude survey 
AGN and stars (short dashed curve; open squares). The ChaMP sources 
have  been reddened to the ChaMPlane field \nh~ values (see text). 
See electronic ApJ for color version.}
\label{Fig4}
\end{figure}

\begin{figure}
\resizebox{\hsize}{!}
{\includegraphics[angle=0]{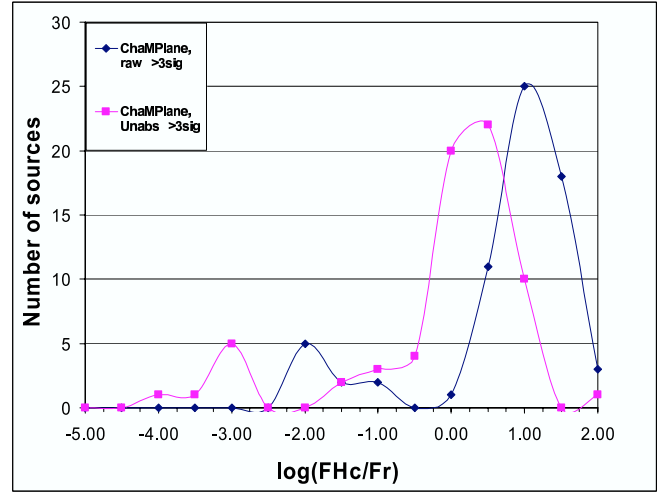}}
\caption{log(\FxFr) distributions (F(Hc) vs. R band)  
for 68 ChaMPlane Anticenter sources (diamonds) with well determined 
fluxes in the Hc band. The flux ratios are de-reddened for the 
full-plane \nh~ (an overestimate, generally) in the comparison 
distribution (squares). Stellar coronal sources are mostly found 
with log(\FxFr) \lsim -2 in this band, and are thus better separated 
from CVs, qLMXBs (and AGN) in the Hc band. See electronic ApJ 
for color version.}
\label{Fig5}
\end{figure}
  
\subsection{Constraints on CV Density or Luminosity Function}
The apparant lack of bright \Halpha~ sources among the ChaMPlane 
Anticenter source IDs allows limits to be derived for the CV 
space density and/or X-ray luminosity function.
We make the conservative assertion that of the 367 sources with  optical
counterparts, at most \about7 could have 
(\Halpha--R) \lsim -0.3 or EW(\Halpha) \gsim 28\AA, which allows 
rejection of dMe stars (see Zhao et al 2005) but not AGN for which 
emission lines can be redshifted in the \Halpha~ filter. 
We include even marginal \Halpha~ objects (all are 
\gsim1.4--2$\sigma$) for a maximum limit, and without spectroscopic 
identification (5 were unobserved and 2 had insufficient counts 
for classification). Since the Sloan Survey CV sample (Szkody et al 2004 
and Papers I and II in this series) find 17\% of their 99 
spectroscopically discovered CVs have EW(\Halpha) \lsim28\AA\, 
our photometric limits for ChaMPlane CVs could be increased 
by this factor (1.17) to \lsim8. Thus a (very) conservative limit 
from the optically identified 
ChaMPlane Anticenter sources is that \lsim2\% are CVs based on these 
\Halpha~ limits. The nearly-half of the Anticenter source sample 
not yet identified optically, with absorbed log(\FxFr) \gsim 1.2  
but for which the unabsorbed limit is \gsim 0,  could allow 
additional CVs, though again the logN-logS results 
suggest these reddened Anticenter sources 
are also dominated by AGN.

We now explore whether these limits, from 14 Anticenter fields, 
provide interesting constraints on the CV number density. Given the 
current estimates from ROSAT (Schwope et al 2002) and 
previously Einstein (Hertz et al 1990) 
that X-ray selected CVs in the solar neighborhood (\lsim 1 kpc)  
are detected with space density \ncv \about 3 \X 10$^{-5}$ \pc3, we 
derive the numbers expected in each of the 14 fields. Given the 
estimated distances of the ROSAT CVs, their luminosities in the 
ROSAT band are peaked at \Lx(0.5-2.5keV) \about 10$^{31}$ \lcgs, with 
only 3 of 49  (Verbunt et al 1997 and Schwope et al 2002) 
between  10$^{29-30}$ \lcgs. 
The 22 optically identified CVs detected by Chandra in the deep 
survey of the globular cluster 47Tuc (Heinke et al 2005) are 
distributed with a cumulative luminosity function described by either a 
power law, N(\gsim F(Sc)) $\propto$ F(Sc)$^{-0.31\pm0.04}$ or 
a lognormal distribution with mean log(\Lx(Sc)) = 31.2$\pm0.32$. 
For a ``typical'' CV spectrum (e.g. a Bremsstrahlung spectrum 
with kT \about5-10 keV), the \Lx(Sc) and \Lx(Hc) values are comparable. 
Thus for \Lx(Hc) = 10$^{30}$ and 10$^{31}$ \lcgs, we derive (Table 2) the 
limiting detection distances d$_{30}$ and d$_{31}$ 
for each field using the (maximum) 
\nh~ values for each as derived from Schlegel et al (1998) 
and given in Table 2. Given the solid angle $\Omega \propto \theta^2$ 
subtended by the full field of the ACIS-I or -S exposure, 
with $\theta$ = 16\arcmin or 8\arcmin, 
respectively (we use the 
full field, and in this simplified treatment do not allow for 
the off-axis decline in sensitivity), we then derive the volume 
in the disk over which CVs could be detected. We assume 
CVs are distributed in the disk with an exponential  
z-distribution (Warner 1995) for which we first assume 
a scale height h = 400pc, or comparable to that 
recently estimated for LMXBs (Grimm et al 2002). We assume the radial 
distribution scale in the disk is much larger 
e.g., \about3.5 kpc, as for stars) and can to first 
order be ignored in this initial treatment. We then  
derive the effective detection volume V$_{eff}$  for each field, 
using  the same formalism adopted by Schwope et al (2002) and  
Tinney et al (1993). Assuming the CVs are distributed 
with a z-distribution as \ncv $\propto$ exp$^{-d (sin b)/h}$, 
the effective detection volume out to distance d is 
\medskip

$V_{eff} = \Omega (h/sin b)^3 (2 - (\chi^2 + 2\chi + 2)e^{-\chi}) $

\medskip

\ni
where $\chi$ = d(sin $b$)/h. We derive this volume for the limiting 
detection distances d$_{30}$ and d$_{31}$ and thus the total number of
CVs,  N$_{CVs}$ = \ncv V$_{eff}$, given in Table 2 as N$_{30}$ 
and N$_{31}$. Summing over the 14 fields gives a 
total of 2.3 CVs expected  
if their characteristic luminosity is \Lx = 10$^{30}$ \lcgs vs. 
53.1 for the \Lx = 10$^{31}$ \lcgs~ value. Changing the assumed 
scale height to 200pc, as assumed by Schwope et al (2002), 
gives corresponding CV numbers 2.0 - 38.3.  
Additional limits are 
given below for a plausible X-ray luminosity function (XLF).  

\begin{figure}
\resizebox{\hsize}{!}
{\includegraphics[angle=0]{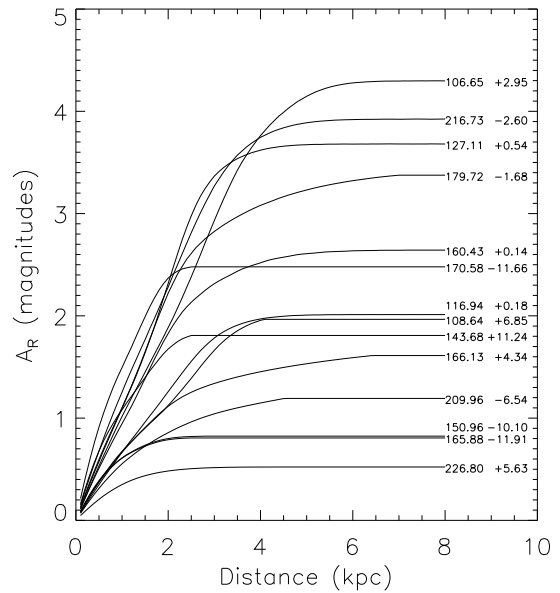}}
\caption{\Ar vs. distance for the 14 Anticenter ChaMPlane 
fields analyzed in Table 1 using the \Av vs. distance model 
of Drimmel et al (2003) for each $l, b$ value shown . 
\Ar = 0.75\Av values are plotted after re-scaling the asymptotic
\nh~ value in the Drimmel model for each field to that given by Schlegel et al
(1998). This re-scaled \Ar vs. d relation is used to estimate 
extinction at distances d$_{30}$ and d$_{31}$ (see text) for each field.}
\label{Fig6}
\end{figure}

We then estimate (Table 2) the range of optical 
R magnitudes these CVs would 
have for the range of log(\FxFv) \about log(\FxFr) 
\about -1 -- +1, typical for CVs 
(Hertz et al 1990, Verbunt et al 1997). However since the optical 
counterparts for the CVs in some of these Anticenter fields must 
be appreciably dimmed by the visual extinction, which in R is A$_R$ =
4.2N$_{22}$, where N$_{22}$ is the \nh~ value in units of 
10$^{22}$ \cmsq as given in Table 2, it is necessary to allow 
for extinction in  
the expected R magnitudes as a function of distance through 
the disk. We do this using the 3D dust model of the Galaxy 
from Drimmel at al (2003). The Drimmel model computes 
\Av vs. d, but for these Anticenter fields  appears to 
underestimate the full-plane 
extinction as indicated by our logN-logS analysis (Hong et al 2005), 
for which the Schlegel et al (1998) full-plane \nh~ gives a better 
match to the extragalactic source number counts in the unabsorbed Hc 
band. Thus we re-normalize the Drimmel \Av values by the full-plane 
\nh~ predicted by Schlegel et al and convert to 
A$_R$ = 0.75A$_V$ for the \Ar vs. d plot in Figure 6. This gives 
\Ar for distances d$_{30}$ and d$_{31}$ and thus the 
expected observed range of R magnitudes for CVs for the \Lx~ 
and \FxFr~ ranges given in Table 2. 

Finally, we estimate the fraction of the predicted X-ray CVs that 
should be identified (``easily'') with magnitudes R \lsim 23 
in our ChaMPlane sample. We use the ROSAT distribution of \FxFv~ as 
plotted in Fig. 4 (and which we assume to be approximately 
the same as the \FxFr~ distribution, given typical CV colors 
(V-R) \about0.3), to derive the cumulative distribution shown 
in Figure 7. This gives the fraction of optical CVs vs. R 
magnitude relative to the R magnitude for log(\FxFr) = +1 since 
these are the {\it faint} optical limits, R$_{31}$ and R$_{30}$ + 5, as 
given in Table 2. The cumulative optical identification fractions, 
C$_{30}$ and C$_{31}$, of the X-ray CVs are read off the plot as 
the fraction value at the ``offset magnitude'' limits R$_{31}$ - 23 
and (R$_{30}$ + 5) -23 and given in Table 2. These fractions are 
then multiplied by the total predicted X-ray CV numbers 
(CV$_{30}$ and CV$_{31}$) to give the 
predicted number of CVs (ID$_{30}$ and ID$_{31}$) 
that should be optically identified  for each field 
at R \lsim23 for these characteristic \Lx values.

\begin{figure}
\resizebox{\hsize}{!}
{\includegraphics[angle=0]{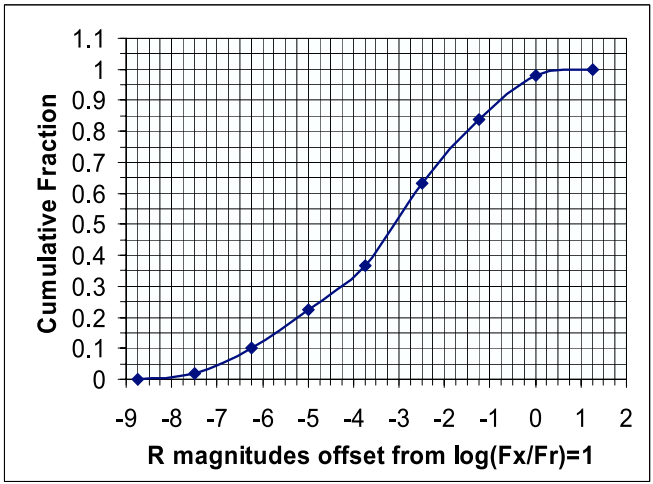}}
\caption{Cumulative distribution of \FxFr~  
for ROSAT CVs (see text) plotted as a function of R magnitude offset, 
for a fixed \Fx, from the optically faint limit at 
log(\FxFr) = +1 for which the R magnitudes expected for CVs in 
each field are given in Table 2. The C$_{30}$ and C$_{31}$ 
values in Table 2 are the fractions of CVs expected with 
R \lsim 23 for CVs with the corresponding \Lx = 10$^{30}$ 
and 10$^{31}$ \lcgs.}
\label{Fig7}
\end{figure}

\section{Conclusions}

The X-ray number counts for ChaMPlane sources in the Anticenter 
(Hong et al 2005) suggest that, as found 
in a moderately deep Chandra Galactic plane survey of $l, b$ 
\about28.5\deg,0\deg~ as well as ASCA Galactic plane 
surveys (Ebisawa et al 2003), at 
fluxes \lsim10$^{-13.5}$ \fcgs the AGN and extragalactic 
source distributions dominate stars and Galactic sources. 
Although the photometry of an ``example'' ChaMPlane field, 
GROJ0422+32 (Zhao et al 2005), shows that 
most of the 40 optical counterparts (out of 62 sources) 
in this field have (V-R) and (R-I) colors 
consistent with their being normal stars, spectroscopy 
(Rogel et al 2005) of 33 of these counterparts showed that 14 
are AGN, 4 are stars and 15 were unidentified (insufficient S/N). 

We have conducted preliminary analysis of 14 Anticenter fields 
to limit their CV content, and contribution from compact binaries 
generally.  Our initial analysis of the  ChaMPlane Anticenter 
CV limits suggests that the 
local space density of CVs may be overestimated by a 
factor of \about3 relative to the 3 \X~ 10$^{-5}$ \pc3 
value derived for the solar neighborhood from the Einstein 
and ROSAT surveys. We derive this as follows. The ROSAT CVs are 
strongly peaked at \Lx \about10$^{31}$ \lcgs, so we first consider the  
limiting case for this \Lx.  
The much larger volumes in the disk (and halo) that
can be reached with Chandra than with ROSAT make it unlikely that 
the  \about38-53 CVs  predicted in the 14 fields could be missed if 
they had local space density \ncv \about3 \X 10$^{-5}$ \pc3 and X-ray
luminosity function with characteristic \Lx \about10$^{31}$ \lcgs. 
Our conservative upper limit of 8 CVs out of 367 optical IDs could 
then allow perhaps 14 CVs for the full sample of 631 sources from 
simple scaling, yielding a deficit factor of \about2.7. 

A stronger limit is obtained from 
the predicted number of CVs which should have magnitudes 
R \lsim23 (``easily'' detected in our Mosaic photometry), given 
our derived \FxFr~ distribution for ROSAT CVs: for CVs 
with characteristic \Lx \about10$^{31}$ \lcgs, some 26 CVs should 
have been optically identified from the ChaMPlane sources. 
This would suggest a factor of \about3 deficit in CV density. Since 
it incorporates ``known'' \FxFr~ flux ratios for CVs and the extinction 
vs. distance model (Drimmel et al 2003) for each field, 
it should be roughly independent of 
the fraction of ChaMPlane sources optically identified.  
ChaMPlane sources with R \lsim23 are consistent with the \FxFr~ 
distributions and extinction for CVs with characteristic 
\Lx \about10$^{31}$ \lcgs~ if \ncv \about1 \X~ 10$^{-5}$ \pc3. 

Allowing for a plausible XLF for disk 
CVs like that recently derived for CVs in the globular 
cluster 47 Tuc (Heinke et al 2005), for which we approximate an 
integral distribution 
N(\gsim\Lx) $\propto $\Lx$^{-0.3}$ for \Lx \lsim10$^{31.5}$ \lcgs, 
steepening to index \about-1.3 for higher luminosities, and 
assuming a low luminosity cutoff at \about10$^{29.5}$ \lcgs, then 
\about0.22 of the sources would be detected with 
\Lx \gsim10$^{31}$ \lcgs. This would appear to suggest a reduction 
by this factor for the CV$_{31}$ total from that in Table 2, or 
comparable to our scaled limit of \lsim14 CVs. However, the extension 
of the XLF to larger \Lx increases the detection volume and thus 
CV number predicted. When combined with the addition of the larger 
number of lower \Lx~ CVs,    
a deficit is still indicated. Full consideration of the XLF 
as well as the the radial scale length of CVs in the galactic disk,   
e.g. \about3.5kpc as for stars,  
will be explored in a more detailed analysis 
with the full ChaMPlane Anticenter sample. 
Our limits are relatively insensitive to the assumed vertical 
scale height (h = 200pc vs. 400pc only reduces the expected 
CV numbers by 28\%) and are consistent with the overall CV 
space density \about10${-5}$ \pc3 for the solar neighborhood 
(Patterson 1998). 

However the Patterson (1998) estimate is dominated (75\%) by 
short-period CVs below the period gap with low \mdot and 
thus \Lx values \lsim10$^{30}$ \lcgs expected. These can be 
detected in these ChaMPlane Anticenter fields out to \about1.5 kpc 
but the numbers expected (CV$_{30}$ in Table 2) are small. 
Additional fields in the Anticenter can test this by 
probing a greater range of $l, b$ values as well as additional regions 
of low \nh. However the best measurements of the low \Lx CV 
population will likely come from our  
low-extinction windows (e.g. Baade's Window) in the Bulge, for which 
analysis is in progress.  Conversely, the luminous 
(\Lx \about 10$^{32-33}$ \lcgs) CV distribution may dominate the 
compact object distributions near the Galactic Center.  
IR imaging of the sources in the cusp around SgrA*
suggests that most are not HMXBs, but are consistent with being 
luminous CVs (Laycock et al 2005). 

With the much larger samples of ChaMPlane data from 
the full survey (these data represent \lsim10\% of the total, and 
cover just 0.6 square degrees), the CV and compact binary 
(qLMXBs, Be-HMXBs) space density and luminosity 
functions will be possible to measure or 
constrain over Galactic scale distances for the first time. The 
ChaMPlane data and images will be available for easy access 
and analysis, as described in the Appendix.  


\bigskip
Many colleagues have contributed to ChaMPlane. We thank P. Green, 
D. Kim, J. Silverman and B. Wilkes for early (and continuing) discussions 
of ChaMPlane vs. ChaMP and D. Kim, especially, for his development of 
the XPIPE processing script. C. Bailyn, A. Cool, P. Edmonds, M. Garcia 
and J. McClintock all provided useful input in the early planning stages, 
and D. Hoard and S. Wachter helped with some early CTIO observations. 
This work is supported in part by NASA/Chandra grants AR1-2001X, AR2-3002A, 
AR3-4002A, AR4-5003A and NSF grant AST-0098683. We thank the Chandra 
X-ray Center for support, and NOAO for support and the Long Term 
Surveys program.  


\appendix

\begin{center}
{ChaMPlane Data Archives and VO ANalysis and Display Tools \\
(see http://hea-www.harvard.edu/ChaMPlane/apj\_appendix.html)}

\end{center}
\medskip

We have developed a suite of analysis 
tools and an on-line database for both the primary ChaMPlane X-ray 
data (source catalogs) as well as the optical counterpart data. 
On-line browse access to ChaMPlane observations and  
data is organized by ObsID (which can also be found on 
the ChaMPlane website, where all observations can be listed 
and sorted) and provides a very 
general and intuitive platform for access to the results. 
In Appendix Fig. 1 
(see http://hea-www.harvard.edu/ChaMPlane/walkthrough\_xray.html) 
we show a screenshot example of access to obsID 676, the 
field originally observed for the BH transient GROJ0422+32. The 
left window allows a browse selection of obsID; the middle window 
enables choice of data return (e.g. formatted text or rdb tables), 
while the right most windows are: top, the source catalog for the 
field selected (showing X-ray source ID number, RA, Dec, etc. of 
the sources in that field); and bottom,  the optically identified 
counterparts for the given source clicked on in the upper window. 
Following the link at the end of this paragraph of text in the 
Appendix website leads to a ``walk-through'' 
tutorial and demonstration of how to create the 
composite screen shown in Fig. 1.

In Chandra cycle 5, we developed and released a Virtual 
Observatory (VO) node 
to facilitate access and on-line analysis of ChaMPlane data.
Database access is via the VO interface in the image 
display and analysis package 
ds9\footnote{http://hea-www.harvard.edu/RD/ds9/},  
which is a major upgrade from the original  
SAOimage package and allows many new features, including VO.  
In general, the user connects through the ds9
toolbar or via any web browser 
(http://hea-www.harvard.edu/ChaMPlane/data/archive). 
Upon connection the ChaMPlane
toolbar is automatically installed on the user's own ds9 running 
locally on the user's machine. 
Commands and results are exchanged between the user's
machine and the ChaMPlane data server at CfA 
running search and analysis scripts. 
This setup has the considerable advantage of
adding custom tools and data to an already familiar platform, 
while maintaining the data archive and analysis and display 
tools at a central facility for maintenance and upgrades. 
A demonstration and link to a ``walk-through''  
of this ChaMPlane VO facility is given in the following paragraph.

The ChaMPlane VO provides access to our 
deep Mosaic optical (V, R, I, \Halpha) images, catalogs of their 
optical photometry, Chandra images with exposure-map
correction, X-ray source data and details of X-ray sources 
with identified optical counterparts.
Images, datatables and plots (e.g. color-magnitude diagrams) 
can be generated for sources detected in 
any desired region on the Chandra or Mosaic images. 
Data can be selected by defining regions with a cursor to overlay 
X-ray or optical source positions selected for a given 
characteristic on either the ACIS or Mosaic field of view. 
An example is shown in Appendix Fig. 2 from the 
source selection shown in Appendix Fig. 1. 
Once again, following the link which is on the website 
Appendix just after Fig. 2 leads to a  ``walk-through'' 
tutorial and demonstration of how to create the 
composite screen shown in web appendix Fig. 2. We note 
that as a matter of scientific interest, the two Chandra sources 
identified as \Halpha~ objects in web appendix Fig. 2 -- namely, the 
two sources with red (\Halpha~ object) source circles and 
optical source ID numbers overplotted in Fig. 2 on both the optical Mosaic 
image (image on left) and the smoothed Chandra image (image on 
right), are identified from WIYN spectroscopy (Rogel et al 2005) 
as a dMe star (optical ID 111135) and a QSO at z = 1.31 (optical ID 
111329), with the (red) wing of MgII at $\lambda$2802 redshifted 
into the \Halpha~ filter! Additional ChaMPlane VO tools for 
interactive inspection of both 
the X-ray and optical data are described on a separate 
link (http://hea-www.harvard.edu/ChaMPlane/walkthrough\_ds9.html)  
at the very end of the web appendix and 
via the Data Archive/Virtual Observatory link 
on the ChaMPlane homepage.

X-ray and optical data are now available on line for the 
initial 14 fields in the Anticenter. X-ray and optical images 
and source catalogs for a given field or region of the 
survey will be added to the database when the first corresponding 
analysis papers are submitted for publication. 
Calibrated V, R, I, \Halpha~ Mosaic images and photometry 
are submitted in parallel to the NOAO Long Term Survey 
archives and CDS; these have already been archived at NOAO for the 
initial 14 Anticenter fields.

{}

\end{document}